\documentclass[12pt]{article}

\usepackage{amssymb,esvect,amsmath,graphicx,latexsym,amsthm,slashed,eso-pic,hyperref}
\usepackage[final]{pdfpages}

\usepackage{epsfig,amsmath,amssymb,verbatim,mathrsfs,hyperref}
\usepackage{xspace}
\usepackage{xcolor}

\usepackage{slashed}
\usepackage{dcolumn}
\usepackage{multirow}
\graphicspath{{./}{Figs/}}
\usepackage{feynmp-auto}

\def\beq{\begin{eqnarray}}
\def\eeq{\end{eqnarray}}
\def\bea{\begin{eqnarray}}
\def\eea{\end{eqnarray}}

\newcommand{\gsim}{\lower.7ex\hbox{$\;\stackrel{\textstyle>}{\sim}\;$}}
\newcommand{\lsim}{\lower.7ex\hbox{$\;\stackrel{\textstyle<}{\sim}\;$}}

\newcommand{\met}{\,/\hspace{-0.25cm}E_T}

\begin{document}
\begin{titlepage}
\hfill{ACFI T19-10}

\noindent
    \vspace{0.2cm}
\begin{center}
  \begin{Large}
    \begin{bf}

Exotic Higgs Decays and the Electroweak  \\ 
\vspace{0.2cm}  Phase Transition 
     \end{bf}
  \end{Large}
\end{center}
\vspace{0.2cm}
\begin{center}
Jonathan Kozaczuk$^{(a,b,c)}$, Michael J.~Ramsey-Musolf$^{(d,c,e)}$, and Jessie Shelton$^{(b)}$
\vspace{1cm}\\
\begin{it}
(a) Department of Physics, University of California, San Diego, La Jolla, CA 92093, USA
\vspace{0.2cm}\\
(b) Department of Physics, University of Illinois, Urbana, IL 61801, USA
\vspace{0.2cm}\\
(c) Amherst Center for Fundamental Interactions, Department of Physics,\\ 
University of Massachusetts, Amherst, MA 01003, USA
\vspace{0.2cm}\\
(d) Tsung-Dao Lee Institute and  School of Physics and Astronomy, \\ Shanghai Jiao Tong University,
800 Dongchuan Road, Shanghai, 200240 China
\vspace{0.2cm}\\
(e) Kellogg Radiation Laboratory, California Institute of Technology, \\
Pasadena, CA 91125 USA
\vspace{0.2cm}\\
email: \emph{\texttt{jkozaczuk@ucsd.edu}},
\emph{\texttt{mjrm@umass.edu}}, 
\emph{\texttt{sheltonj@illinois.edu}}
\vspace{0.2cm}
\end{it}
\end{center}
\center{\today}

\begin{abstract}
Light new physics weakly coupled to the Higgs can induce a strong first-order electroweak phase transition (EWPT). Here, we argue that scenarios in which the EWPT is driven first-order by a light scalar with mass between $\sim 10$ GeV --  $m_h/2$ and small mixing with the Higgs will be conclusively probed by the high-luminosity LHC and future Higgs factories. Our arguments are based on analytic and numerical studies of the finite-temperature effective potential and provide a well-motivated target for exotic Higgs decay searches at the LHC and future lepton colliders.

\end{abstract}
\end{titlepage}
\section{Introduction}

Determining the thermal history of electroweak symmetry breaking (EWSB) in the early Universe is an important challenge for particle physics and cosmology. In principle, our Universe could have started out in a phase with broken electroweak symmetry, either due to a low reheat temperature after inflation, or from symmetry non-restoration effects~\cite{Weinberg:1974hy, Meade:2018saz}. Another option, predicted in the Standard Model (SM)~\cite{Kajantie:1995kf,Kajantie:1996mn,Kajantie:1996qd,Csikor:1998eu}, is that electroweak symmetry was broken during a thermodynamic cross-over  transition. It is interesting to ask how the SM picture would be changed in the presence of physics beyond the Standard Model (BSM). Perhaps the most intriguing possibility is that electroweak symmetry breaking occurred via a first-order phase transition (PT). This last possibility has attracted considerable attention since it could provide one of the necessary conditions for electroweak baryogenesis (for a review, see~\cite{Morrissey:2012db}), as well as produce a stochastic gravitational wave (GW) background observable at future experiments such as LISA~\cite{Caprini:2015zlo, Caprini:2019egz}. Our current understanding of the Higgs sector is not sufficiently comprehensive to distinguish between these qualitatively different scenarios. There has therefore been a substantial recent effort to suggest new ways in which to probe the nature of the electroweak phase transition (EWPT) with collider searches and GW interferometers (see e.g.~\cite{Caprini:2015zlo, Caprini:2019egz, Cepeda:2019klc,Abada:2019lih,CEPCStudyGroup:2018ghi,DiMicco:2019ngk} and references therein).  Most of this work has concentrated on scenarios wherein the BSM mass scale lies at or above the electroweak scale. In contrast to BSM models intended to address other open problems in particle physics and cosmology, such as dark matter or the origin of neutrino masses, the mass scale associated with interactions driving a first-order EWPT cannot be arbitrarily heavy with respect to the weak scale. Thus, uncovering a departure from the SM thermal history of EWSB provides a compelling target for experiment \cite{MRM19}.

An intriguing, alternate possibility is that {\em light} BSM particles (with masses {\it e.g.}, below $m_Z$) coupled to the Higgs could result in a strongly first-order EWPT (SFOEWPT). These scenarios are also appealing experimentally, though they provide qualitatively distinct challenges compared to the case where all BSM particles are heavier than the SM Higgs. In this paper we address the experimental signatures associated with a SFOEWPT catalyzed by a light new degree of freedom. Given the stringent constraints set by LEP on new light  particles carrying electroweak charge, as well as LHC limits on low-mass strongly-interacting states, we consider new particles 
transforming as singlets under the SM gauge group. In adding a single species with renormalizable interactions with the Higgs boson, one can imagine at least two possibilities:  new gauge singlet fermions interacting with the Higgs and lepton doublet(s) through the so-called ``neutrino portal'', or singlet scalar fields coupled to the Higgs through new terms in the scalar potential. Fermions can drive the EWPT first-order through loop effects when they achieve a large mass across the transition~\cite{Carena:2004ha}, and so it is difficult to envision a case where light singlet fermions can accomplish this while remaining consistent with experimental constraints. Singlet scalar fields, on the other hand, can influence the EWPT at tree-level, even with small couplings to the Higgs. In what follows we therefore focus on cases in which a light singlet scalar degree of freedom catalyzes a SFOEWPT.  The relevance of this regime for a SFOEWPT was previously identified in Ref.~\cite{Profumo:2007wc}, where it was observed that the corresponding exotic Higgs decay branching ratio could be significant.  After the discovery of the SM-like Higgs, studies of phase transitions in singlet extensions of the SM have primarily focused on the regime where the singlet mass $m_s$ exceeds half the Higgs mass (see e.g.~\cite{No:2013wsa, Curtin:2014jma, Profumo:2014opa, Huang:2016cjm, Kotwal:2016tex, Curtin:2016urg, Chala:2016ykx, Beniwal:2017eik, Kurup:2017dzf, Chen:2017qcz, Huang:2017jws, Li:2019tfd} and references therein).  Here, we consider the regime where $m_s < m_h/2$, with the aim of exploring what exotic Higgs decay searches are revealing about the possible thermal histories of the early universe. 

In what follows, we argue that a SFOEWPT catalyzed by a singlet scalar lighter than half the Higgs mass is strongly correlated with the branching ratio of the 125 GeV SM-like Higgs to pairs of the lighter scalar when the singlet-Higgs mixing angle is small (as motivated by current experimental limits). Using simple (semi-)analytical arguments and numerical scans of the parameter space, we motivate targets for exotic Higgs decay searches and investigate the extent to which such searches  can probe the nature of the electroweak phase transition. We find that
\begin{itemize}
\item Simple theoretical arguments suggest a lower bound on the exotic Higgs decay branching ratio consistent with a light scalar-catalyzed SFOEWPT, for which we provide approximate semi-analytical expressions.
\item Experimentally, measurements of Higgs boson properties at the LHC already imply significant limits on this scenario. Our numerical scans do not find any surviving parameter space accommodating a SFOEWPT induced by a light scalar with mass between $\sim 30$ GeV and $m_h/2$.  However, viable parameter space still exists for lighter scalars.
\item The High Luminosity LHC (HL-LHC) and prospective future lepton colliders will be able to improve this sensitivity down to $\sim 10$ GeV in the channels we consider, and potentially further. 
\end{itemize} 

While these results come with certain caveats (discussed below), the work presented here nevertheless provides an important physics target for current and future exotic Higgs decay searches. We will focus on light scalars with masses above the $b$-quark mass, although it would also be interesting to extend our results to lower masses in the future. 

The remainder of this study is structured as follows. In Sec.~\ref{sec:pot} we discuss models with an additional light singlet scalar coupled to the Higgs, describing the phenomenology at both zero  and finite temperature. In Sec.~\ref{sec:bounds} we put forward simple semi-analytic arguments to motivate a lower bound on the branching ratio of the Higgs to two light scalars when requiring a strong first-order EWPT, and perform numerical scans of the parameter space that support the results from the analytic treatment. In Sec.~\ref{sec:brexo}, we analyze the corresponding implications for exotic Higgs decay searches at colliders. In Sec.~\ref{sec:conc} we discuss some caveats to our arguments, and conclude.

\section{Light Scalars Coupled to the Higgs} \label{sec:pot}

We are interested in the electroweak phase transition in models with a relatively light real singlet scalar, $S$, coupled to the Standard Model-like Higgs field, $H$.  The scalar potential of interest consists of the Standard Model Higgs potential augmented by all possible renormalizable terms involving $S$:
\beq
\begin{aligned} \label{eq:pot}
V=&-\mu^2 \left|H\right|^2 + \lambda \left|H\right|^4 +\frac{1}{2} a_1 \left|H\right|^2 S + \frac{1}{2} a_2 \left|H\right|^2 S^2 + b_1 S + \frac{1}{2}b_2 S^2 +\frac{1}{3} b_3 S^3+\frac{1}{4}b_4 S^4,
\end{aligned}
\eeq
where we follow the notation of Ref.~\cite{Profumo:2007wc}.
After electroweak symmetry breaking and at zero temperature, we can parametrize the fields (in unitary gauge) as
\beq
H = \frac{1}{\sqrt{2}}\left( \begin{array}{c}
0 \\
v+h
\end{array}
\right),
\hspace{0.3 cm} S =\left( v_s+ s \right)
\eeq
where $v=246$ GeV is the vacuum expectation value (VEV) of the Higgs, and $v_s$ is the singlet VEV. Since $S$ is a gauge singlet and only couples to $H$ in the Standard Model, one is free to shift $S$ by a constant without changing the physical predictions of the theory~\cite{Profumo:2007wc, Espinosa:2011ax}. This shift is often used to either remove the tadpole term proportional to $b_1$, or to set $v_s=0$. We choose the latter. 

In the absence of additional symmetries, the two fields $h$ and $s$ will generally mix. The mass eigenstates can be parametrized as 
\beq
\begin{aligned}
&h_1 = h \cos \theta + s \sin\theta \\
&h_2 = -h \sin \theta + s \cos\theta, \label{eq:mix1}
\end{aligned}
\eeq
with corresponding masses $m_1$ and $m_2$. The eigenstates are ordered by mass, such that $m_1 \leq m_2$, while the physical range of $\theta$ is $\pi/4 < \theta< 3 \pi/4$. Since we are interested in exotic Higgs decay signatures of the electroweak phase transition, we will take $h_1$ to be mostly singlet-like, and $h_2$ to correspond to the Standard Model-like Higgs with $m_2 = 125$ GeV. In terms of the mass eigenstates, the trilinear scalar interactions can be parametrized as
\beq
V \supset \frac{1}{6}\lambda_{111} h_1^3 + \frac{1}{2}\lambda_{211}h_2 h_1^2 + \frac{1}{2}\lambda_{221}h_2^2 h_1 + \frac{1}{6}\lambda_{222} h_2^3.
\eeq
Exotic Higgs decays $h_2 \to h_1 h_1$ are governed by $\lambda_{211}$, with partial width given by
\beq \label{eq:decay}
\Gamma(h_2 \to h_1 h_1) = \frac{1}{32 \pi^2 m_2} \lambda_{211}^2 \sqrt{1-\frac{4 m_1^2}{m_2^2}}.
\eeq
We are thus interested in the correlation between the strength of the electroweak phase transition and $\lambda_{211}$. 

Several of the parameters in Eq.~(\ref{eq:pot}) can be replaced by more physical quantities. Requiring  $m_2 = 125$ GeV, $v=246$ GeV,  $v_s=0$ and using $m_1$ and $\theta$ as input parameters, one finds 
\beq
\begin{aligned} \label{eq:params}
& \mu^2 = \lambda v^2, \quad \lambda = \frac{1}{2 v^2} \left( m_1^2 \cos^2 \theta + m_2^2 \sin^2 \theta \right) \\
b_1 = - \frac{1}{4} a_1 v^2, \quad &b_2 = -\frac{1}{2} a_2 v^2 + m_2^2 \cos^2 \theta + m_1^2 \sin^2\theta, \quad a_1 = \frac{1}{v} \left(m_2^2-m_1^2\right) \sin 2\theta.
\end{aligned}
\eeq
We will therefore take $m_1$, $\cos\theta$, $a_2$, $b_3$ and $b_4$ as our input parameters. Note that $b_4>0$ is required for the stability of the potential.

Further simplifications arise from noting that LEP constrains $|\cos\theta|$ to be quite small for light scalars. 
Currently, $|\cos \theta| \lesssim 0.07$ is allowed for all masses between 10 GeV $\leq m_1 \leq 120$ GeV~\cite{Robens:2015gla}, with larger mixing angles allowed at large masses.  Although the precise limits vary by mass, ongoing precision Higgs analyses at the LHC and, potentially, at future colliders will continue to probe smaller and smaller values of $|\cos\theta|$, likely down to the $\mathcal{O}(0.01)$ level~\cite{Draper:2018ljh}.  We will focus here on the experimentally challenging regime with  $|\cos\theta|\lesssim 0.01$, where we will see that exotic Higgs decays provide a powerful and complementary probe.  This small-mixing regime has the added benefit of simplifying our analytic analysis below. Note that mixing angles in this range typically require at least percent-level fine-tuning to realize, unless they correspond to a small breaking of an underlying symmetry. 

A particularly interesting limit of the light scalar scenario arises when one imposes a $Z_2$ symmetry under which $S\to -S$ but all other fields are neutral. We will refer to this as the $Z_2$ limit. In the parametrization introduced above, the $Z_2$ limit corresponds to the general model with $b_3=0$ and $\cos\theta=0$ (and therefore $a_1, b_1=0$).
The singlet does not mix with the Higgs in this limit, and, if produced at a collider, would escape the detector as missing energy. If the $Z_2$ symmetry is spontaneously broken, the model is described by the more general potential in Eq.~(\ref{eq:pot}) with specific relationships among the parameters. We do not specifically consider the spontaneously broken $Z_2$ scenario further, since it corresponds to a specific subspace of the parameter space of the full singlet model, and  in the small mixing angle regime is thus also covered by our analysis.  However, it is worth bearing in mind that in the non-$Z_2$ symmetric case, $\cos\theta$ and $a_2$ are independent parameters, meaning that a given exotic Higgs branching ratio can be realized for a range of possible $\cos\theta$, while in models with a spontaneously broken $Z_2$, the mixing angle cannot be dialed separately from the exotic branching fraction. 

In the small-$|\cos\theta|$ regime, we can expand quantities around their $|\cos\theta|=0$ values, with higher-order corrections in $|\cos\theta|$ being highly suppressed. Defining $\theta^\prime \equiv \pi/2-\theta$, we have
\beq
\label{eq:params_approx}\begin{aligned}
\lambda = \frac{1}{2v^2} &m_2^2 + \mathcal{O}(\theta^{\prime2}), \quad b_2 = -\frac{1}{2} a_2 v^2 + m_1^2 + \mathcal{O}(\theta^{\prime2}) \times m_2^2,\\
 a_1 &= 0 + \mathcal{O}(\theta^\prime) \times \frac{m_2^2}{v}, \quad b_1 = 0 + \mathcal{O}(\theta^\prime) \times m_2^2 v.
\end{aligned}
\eeq
We will make use of these expansions in our analytic arguments below, although we use the full expressions of Eq.~(\ref{eq:params}) for our numerical scans. In the small $|\cos\theta|$ limit, the coupling controlling the $h_2\to h_1 h_1$  branching ratio becomes simply 
\beq \label{eq:L211}
\frac{\lambda_{211}}{v} = -a_2  + \mathcal{O}(\theta^{\prime2}) 
\eeq
(for a full expression see e.g.~\cite{Profumo:2007wc, Chen:2014ask}). 
As we argue below, requiring a strong first-order EWPT implies a preferred range for $a_2$, and is therefore tightly correlated with a minimum branching ratio for exotic Higgs decays. To flesh out this connection, we need to consider the theory at finite temperature, which we do using an effective potential approach, described in the following subsection.

\subsection{Finite Temperature}

At finite temperature, the relevant quantity to analyze is the free energy, also known as the finite-temperature effective potential, $V_{\rm eff}$. The leading finite-temperature corrections to Eq.~(\ref{eq:pot}) can be written as
\begin{align}
\label{eq:VT}
 \Delta V^T  =\frac{T^4}{2\pi^2}\left[\sum_i \pm n_i J_{\pm}\left(\frac{m^2_i(h,s)}{T^2}\right)\right] ,
 \end{align}
where  $h$, $s$ should be understood as the Higgs and singlet background field values, respectively, and
  \begin{equation}
  J_{\pm}(x)= \int_0^\infty dy\, y^2\log\left[1\mp \exp(-\sqrt{x^2+y^2})\right] .
 \end{equation}
The sum in Eq.~(\ref{eq:VT}) runs over all species coupled to the scalar fields, with the upper (lower) sign corresponding to bosons (fermions), respectively. When $m/T$ is small, the thermal functions $J_\pm$ take on particularly simple forms:
\beq
\begin{aligned}
 T^4 J_-\left(\frac{m^2}{T^2}\right)= &\frac{7\pi^4 T^4}{360}-\frac{\pi^2m^2 T^2}{24}-\frac{(m^4)}{32}\log\frac{m^2}{a_f T^2}, \\
 T^4 J_+\left(\frac{m^2}{T^2}\right) =&-\frac{\pi^4 T^4}{45}+\frac{\pi^2m^2 T^2}{12}-\frac{T\pi (m^2)^{3/2}}{6}-\frac{(m^4)}{32}\log\frac{m^2}{a_b T^2}, \\
\end{aligned}
\eeq
where $a_f$ and $a_b$ are numerical constants (see e.g.~Ref.~\cite{Quiros:1999jp} for more detail). In practice, these ``high-temperature'' expansions are good approximations to  $J_\pm$ for $m/T \lesssim 2$. For the relatively light singlet scalars of interest here, this high-temperature expansion is well-justified. 

In a power-counting scheme where all masses are parametrically\footnote{In certain cases relying on relatively large couplings to impact the EWPT, other power counting schemes, along the lines of e.g.~\cite{Katz:2015uja}, can be used to modify the arguments that follow.} $m_i\sim g \phi$, where $\phi=h,s$,
the high-temperature expansion to $\mathcal{O}(g^2)$ involves adding terms constant and quadratic in $m(h,s)$  to the tree-level, zero-temperature potential in Eq.~(\ref{eq:pot}).  The resulting potential is a gauge-invariant $\mathcal{O}(g^2)$ approximation to the full finite-temperature effective potential. This is the approximation we will focus on in this paper for analytical results. Keeping terms to $\mathcal{O}(g^3)$ amounts to  adding in addition a non-analytic cubic term arising from the bosonic degrees of freedom. This contribution to the effective potential is gauge-dependent (see e.g.~\cite{Patel:2011th} for a detailed discussion of this issue). We perform numerical scans including the leading $\mathcal{O}(g^3)$ terms (from the gauge bosons) and explicitly verify that including them in Landau gauge does not appreciably change our results.
Of course, fixed-order perturbation theory breaks down in the symmetric phase owing to the appearance of new massless modes in the spectrum \cite{Parwani:1991gq,Carrington:1991hz,Arnold:1992rz}, and so our arguments here should be understood as a preliminary guide. 

\subsection{Radiative corrections at zero temperature}

There are also zero-temperature radiative corrections to the effective potential. At one loop, they are given by the Coleman-Weinberg contribution,
\beq
\Delta V_{\rm 1-loop} = \sum_i \frac{\pm n_i}{64 \pi^2} m_i^4(h,s)\left[\log\left(\frac{m_i^2(h,s)}{\mu_R^2}\right) - c_i \right] ,
\eeq
correcting the scalar potential in Eq.~(\ref{eq:pot}) so that $V_{\rm eff} = V+\Delta V^T+\Delta V_{\rm 1-loop}$. Here the upper (lower) sign is for bosons (fermions), the $c_i$ are renormalization group scheme-dependent constants, and $\mu_R$ is the renormalization scale. Importantly, the Coleman-Weinberg potential depends on the field-dependent masses to the fourth power. Again parametrically taking $m_i\sim g \phi$, the zero-temperature radiative corrections are formally of $\mathcal{O}(g^4)$ or higher, and so are subdominant to thermal corrections which arise first at $\mathcal{O}(g^2)$.  For the small couplings and masses in the regime of interest, the largest 1-loop corrections will come from the top quark, effectively modifying e.g.~the quartic coupling $\lambda$ at the $\sim 1-10\%$ level. As explained below, a barrier separating phases can already arise at $\mathcal{O}(g^2)$ in our scenario, and since we will be interested in small couplings, these 1-loop $\mathcal{O}(g^4)$ effects should not dramatically impact our results, which focus on the smallest couplings consistent with a SFOEWPT. In what follows, we will therefore not consider the zero-temperature Coleman-Weinberg corrections.  Nevertheless, it would be worthwhile to extend our analysis to $\mathcal{O}(g^4)$ in the future to explicitly verify the applicability of the power-counting arguments above.

\section{The electroweak phase transition and the Higgs couplings to light scalars} 
\label{sec:bounds}

Given the finite-temperature effective potential, we can now consider the parameter space of the theory accommodating a strong first-order electroweak phase transition. We define a strong first-order electroweak phase transition by the criterion
\beq
\label{eq:sfocond}
\frac{v_c}{T_c} \geq 1
\eeq
where $v_c$ is the value of the Higgs background field at the critical temperature, $T_c$, at which the high- and low-temperature phases are degenerate. Note that in the $\mathcal{O}(g^2)$ approximation, $v_c/T_c$ is gauge-invariant, while it is not once higher-order $\mathcal{O}(g^3)$ corrections are included (see~\cite{Patel:2011th}). In addition, as the phase transition proceeds via bubble nucleation, the corresponding nucleation rate must be sufficiently large to enable the transition to complete. This nucleation requirement is distinct from the criterion in Eq.~(\ref{eq:sfocond}).

For small $|\cos \theta|$, as required by current experimental constraints, the leading coupling of the singlet scalar to the Higgs is $a_2$. In the absence of any couplings between $S$ and $H$, the electroweak phase transition should proceed as in the Standard Model, namely as a cross-over. Therefore, for $S$ to catalyze a strongly first-order electroweak phase transition, we expect that $a_2$ cannot be too small, and, correspondingly, neither can $\lambda_{211}$. Requiring a strongly first-order EWPT can therefore set a concrete target for exotic Higgs decay searches, which are sensitive precisely to $\lambda_{211}$. 

Below, we put forward semi-analytical arguments for a lower bound on $a_2$, and hence exotic Higgs branching ratios, from the requirement that a SFOEWPT occurs.  We work in a strict small-$|\cos \theta|$ limit so that the expressions in Eq.~(\ref{eq:params_approx}) can be used; as we will see, the lower bound obtained from this analysis is self-consistently within the regime of validity of the expansion. For simplicity, we also make use of the high-temperature approximation ($m/T \ll 1$) up to $\mathcal{O}(g^2)$, i.e.~keeping only the (gauge-independent) $T^2$ corrections. We then confirm our reasoning with a numerical analysis of the phase structure across the relevant parameter space in a high-temperature approximation up to $\mathcal{O}(g^2)$ and $\mathcal{O}(g^3)$,
 retaining all dependence on $\cos \theta$ and performing a full numerical tunneling calculation.

\subsection{Semi-analytic arguments in the small-mixing limit}
\label{sec:analytic}

There are several conditions that must be satisfied by the finite-temperature effective potential for a strongly first-order EWPT to be possible without violating phenomenological constraints.
We will outline these requirements and show how they can be combined to obtain a lower bound on $a_2$ in the small mixing angle regime, for a given value of $m_1$.

To begin, we will focus on transitions between minima separated by a tree-level barrier. These are so-called ``two-step transitions'', for which the symmetry breaking pattern is approximately
\beq\label{eq:pattern}
 (h=0, s\simeq 0) \to (h=0, s\neq 0) \to (h\neq 0, s \simeq 0)
\eeq
where $h$ and $s$ denote the Higgs and singlet background fields at finite temperature~\cite{Profumo:2007wc, Espinosa:2011ax, Curtin:2014jma, Xiao:2015tja}.  At high temperatures, the interactions proportional to $b_1$, $a_1$ and $b_3$ contribute an additional singlet tadpole term to the finite-temperature effective potential, and hence can generate a VEV for the singlet (recall at that $T=0$ this tadpole term removes the singlet vev). However, $b_1$ and $a_1$ are suppressed by $|\cos\theta|$, and the finite-temperature tadpole term arising primarily from $b_3$ is typically numerically small~\cite{Profumo:2007wc}, which is why for small $|\cos\theta|$ the first step of the transition typically proceeds from a $(h=0, s\simeq 0)$ minimum.

As discussed above, we will treat the finite-temperature effective potential in a high-temperature expansion and only keep terms to $\mathcal{O}(g^2)$. This means that we do not include the finite-temperature cubic term in $V_{\rm eff}$ or the Coleman-Weinberg corrections. Neglecting the Coleman-Weinberg contribution to the effective potential is a reasonable approximation especially for light singlets, since these corrections are suppressed by the loop factor $\sim 1/16\pi^2$ and are formally of $\mathcal{O}(g^4)$, whereas the Higgs-singlet interactions driving the first-order transition generically occur at tree-level and $\mathcal{O}(g^2)$ (we discuss possible exceptions in Sec.~\ref{sec:flat}). Furthermore, the high-$T$ expansion is also justified, since we expect $m_i/T$ for all species to be small near the broken phase for both $h$ and $s$ (this condition can be checked \textit{a posteriori} and we indeed find this to be the case at the critical temperature for the small values of $a_2$ relevant for our lower bound). 

In keeping only terms up to $\mathcal{O}(g^2)$ in the finite temperature effective potential, Eq.~(\ref{eq:pattern}) is in fact the only viable EWSB pattern accommodating a SFOEWPT, since without a ``singlet'' ($s\neq 0$, $h=0$) minimum at finite temperature, $\mathcal{O}(g^3)$ corrections to the effective potential are required to produce a barrier between phases. Considering the finite-temperature effective potential to $\mathcal{O}(g^2)$ and dropping all terms proportional to $\theta^\prime$, we find that such singlet minima occur at
\beq \label{eq:s_minima}
s(T) \simeq - \frac{b_3 \pm \sqrt{b_3^2 - 4 b_4 b_2(T)}}{b_4}
\eeq
where $b_2(T)$ is the finite-temperature singlet mass term,
\beq
b_2(T) = b_2 + \beta T^2 \ ,
\eeq
with 
\beq
\beta \equiv \frac{1}{12}\left(2 a_2+3 b_4\right)\ .
\eeq
The upper (lower) sign in Eq.~(\ref{eq:s_minima}) corresponds to $b_3>0$ ($b_3<0$). 
In what follows, quantities written without a temperature argument refer to their $T=0$ values. In analyzing the $T>0$ behavior of the theory, it is useful to refer to the ``electroweak temperature", $T_{\rm EW}$, at which the coefficient of the Higgs quadratic term changes sign:
\beq
T_{\rm EW} \simeq m_2/\sqrt{2 \alpha},
\eeq
where we have defined
\beq
\alpha \equiv \frac{1}{48}\left(24 a_2 + 9 g^2 + 3 g^{\prime 2}+ 24 \lambda + 12 y_t^2\right)
\eeq
such that $\mu^2(T) = \mu^2 - \alpha T^2$. For small $a_2$, $T_{\rm EW}\approx  140$ GeV, as in the Standard Model.  Meanwhile, the criterion of Eq.~(\ref{eq:sfocond}) for a SFOEWPT requires the Higgs quadratic term to be sufficiently negative at the critical temperature $T_c$ to support a VEV of the requisite size, 
\beq
\mu^2 (T_c) \geq \lambda T_c^2.
\eeq
This inequality implies that a SFOEWPT in the singlet model requires $T_c < T_{\rm EW}$. 

We now consider a number of conditions needed for the occurrence of the two-step trajectory outlined above.

\noindent \textbf{1.} A necessary (but not sufficient) condition for the existence of the $s\neq 0$ minimum at $T\geq 0$ is 
\beq \label{eq:existence}
a_2 > \frac{1}{v^2}\left(2 m_1^2 - \frac{b_3^2}{2 b_4} \right).
\eeq
Note that for $b_3 =0$, as in the $Z_2$ limit, this requirement already places a rather stringent lower bound on $a_2$ consistent with a strong first-order EWPT:
\beq \label{eq:a2_bound_Z2}
\boxed{
a_2 > \frac{2 m_1^2}{v^2} \quad (Z_2 \, {\rm limit})
}
\eeq
(see also Ref.~\cite{Curtin:2014jma}). This requirement is simply the statement that $b_2<0$. Away from the $Z_2$ limit, non-zero $b_3$  and a small  value of $b_4$ can relax this lower bound on $a_2$, \textit{i.e.}~allow for a strong first-order EWPT with positive values of $b_2$. In that case, several additional requirements, described below, must be imposed to obtain a lower bound on $a_2$. In the rest of this discussion, we therefore concentrate on $b_2>0$.

\begin{figure}[!t]
\begin{center}
\includegraphics[width=0.49\textwidth]{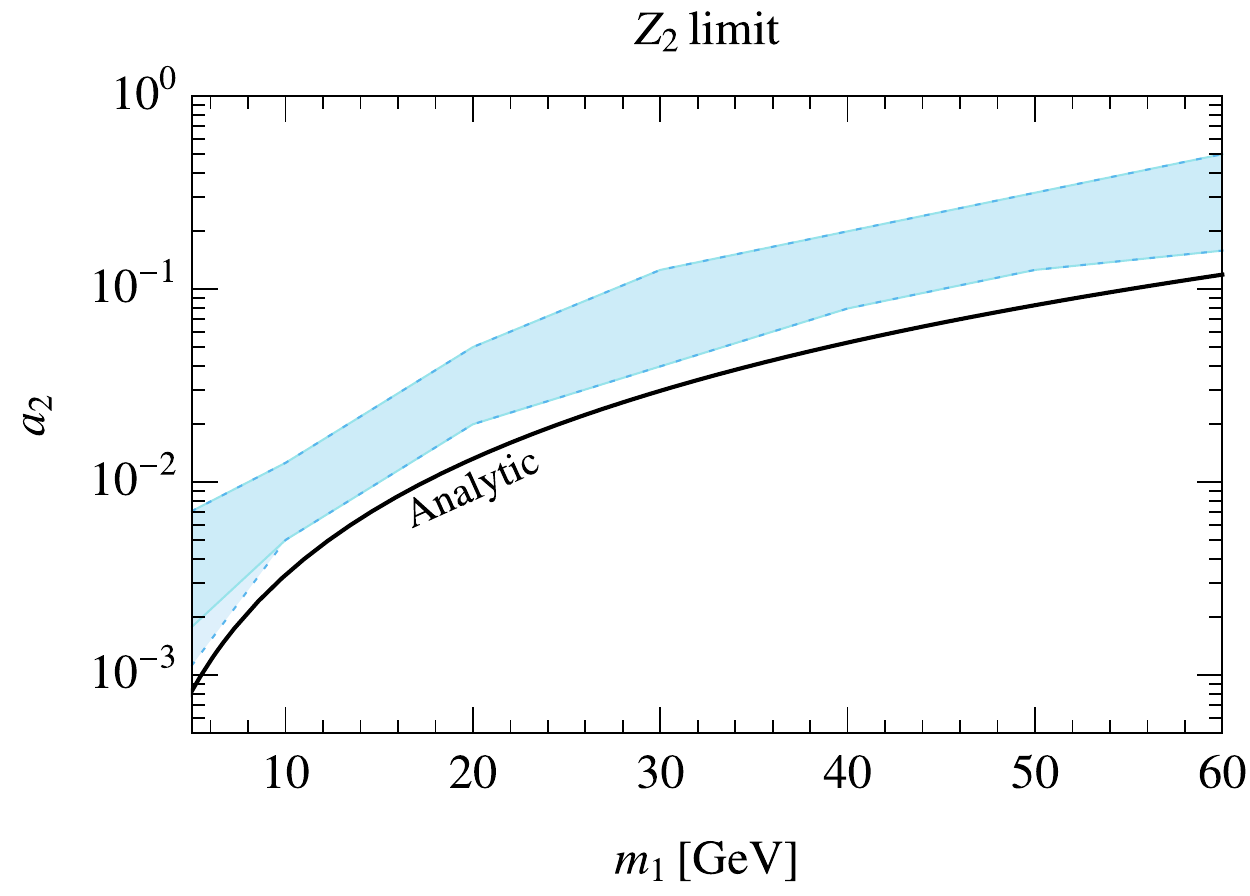}
\caption{Values of the coupling $a_2$ consistent with a strongly first-order electroweak phase transition in the $Z_2$ limit. The black curve corresponds to the analytic lower bound of Eq.~(\ref{eq:a2_bound_Z2}). The dark shaded region shows the results from a numerical scan of the parameter space using the high-temperature effective potential to $\mathcal{O}(g^2)$ and including a calculation of the tunneling rate, as discussed in Sec.~\ref{sec:numerics}. The lighter shaded region bounded by the dotted contours indicates the results of numerical scans including the leading $\mathcal{O}(g^3)$ corrections in Landau gauge (note that this region coincides with the $\mathcal{O}(g^2)$ results across most of the parameter space).
}
\label{fig:res_Z2} 
\end{center}
\end{figure}

\noindent \textbf{2.} The electroweak vacuum must be the global minimum of the potential at zero temperature\footnote{Strictly speaking, the electroweak vacuum need not be the global minimum of the potential at zero temperature. It can instead be metastable with a lifetime longer than the current age of the Universe. For simplicity, however, we require absolute stability.}. For non-zero $b_3$, plugging Eq.~(\ref{eq:s_minima}) into the expression for the potential and requiring $V(0,s) > V(v,0)$ yields
\beq
\label{eq:s_stable}
\left(1+\sqrt{1-\xi_0}\right)^2 f(\xi_0)b_3^4 <\frac{48\mu^4 b_4^3}{\lambda} \ ,
\eeq
where we have defined the quantity 
\beq
\xi_T\equiv \frac{4b_4 b_2(T)}{b_3^2} \,
\eeq
indicating temperature dependence with a subscript, and
\beq
f(\xi)\equiv 2+2\sqrt{1-\xi}-3\xi.
\eeq
Note that restricting ourselves to $b_2>0$ corresponds to $\xi_0>0$.

\noindent \textbf{3.}  The $s\neq 0$ extremum should become a minimum of the potential at some temperature $T_s \geq T_{\rm EW}$, so that the field can transition to this singlet minimum before the origin is destabilized at $T_{\rm EW}$.  For $T<T_{s}$, including the temperature $T_*$ at which the phase transition completes (defined precisely in Sec.~\ref{sec:numerics}), this requirement yields the condition
\beq
\label{eq:s_stable2}
\mu^2(T) < \frac{a_2 b_3^2}{8 b_4^2} \left(1+\sqrt{1-\xi}\right)^2 ,
\eeq
where $\mu^2(T)$ is the finite-temperature Higgs mass-squared term. Note that this condition implies 
\beq
a_2>0
\eeq
for two-step transitions with a tree-level barrier.

\noindent \textbf{4.} The $s\neq 0$ minimum must have lower free energy than the origin at some temperature above $T_{\rm EW}$, yielding the condition
\beq \label{eq:b3_lower}
\xi_{\rm EW} < \frac{8}{9}.
\eeq
Taken as a requirement on $|b_3|$, Eq.~(\ref{eq:b3_lower}) is more stringent than Eq.~(\ref{eq:existence}) and, when inserted into Eq.~(\ref{eq:s_stable}), further constrains the possible values of $b_4$ and $a_2$ allowed by the physical requirements discussed above.

\noindent \textbf{5.} The singlet vacuum must have higher free energy than the electroweak vacuum for temperatures in the range between  the critical temperature $T_{c}$ and the temperature at which the phase transition completes, $T_*$.  This condition yields an inequality of the form Eq.~(\ref{eq:s_stable}), but where $\xi_0\to \xi_T$ and $\mu^2\to\mu^2(T)$, 
yielding
\beq\label{eq:higher_smin}
\left(1+\sqrt{1-\xi_*}\right)^2 f(\xi_*)b_3^4 <\frac{48\mu(T_*)^4  b_4^3}{\lambda }.
\eeq
 Substituting Eq.~(\ref{eq:s_stable2}) into the RHS of Eq.~(\ref{eq:higher_smin}) then leads to a requirement on $a_2$
\beq
\label{eq:a2_bound1}
a_2^2 > \frac{4 b_4 \lambda }{3\left(1+\sqrt{1-\xi_*}\right)^2} f(\xi_*).
\eeq
However, note that this requirement alone is still not sufficient to derive a lower bound on $a_2$. The function $f(\xi)\to 0$ as $\xi\to 8/9$, implying that when the transition occurs near $T_{\rm EW}$ the parameter $a_2$ could be arbitrarily small.

\noindent \textbf{6.} We now consider the nucleation requirement. At the temperature $T_*$ at which tunneling completes, the potential must be away from the so-called ``thin-wall limit''. The thin-wall regime describes cases where the splitting between the minima of the potential is significantly smaller than the height of the barrier separating the two phases. The tunneling rate is highly suppressed in this regime, preventing successful completion of the phase transition (see e.g.~\cite{Linde:2005ht} for a pedagogical discussion of this point). Denoting the singlet minimum as $\phi_s$, the broken phase minimum as $\phi_h$, and the location of the barrier peak along the correct tunneling trajectory as $\phi_b$, we require
\beq
\label{eq:tunneling}
\frac{V(\phi_s, T_*)-V(\phi_h,T_*)}{V(\phi_b,T_*)-V(\phi_h,T_*)} > \Delta
\eeq
where $\mathcal{O}(0.1) \lesssim \Delta<1$ is a number that can be chosen empirically to be consistent with the results of our numerical scans, described below. Note that Eq.~(\ref{eq:tunneling}) should simply be viewed as an approximate necessary (but not sufficient) condition for the completion of the phase transition, and should not be expected to reflect the full impact of requiring successful tunneling.  The full tunneling requirement generally features a complicated dependence on the masses and couplings, and will be discussed in detail in Sec.~\ref{sec:numerics}.  Nevertheless, Eq.~(\ref{eq:tunneling}) will provide physical insight into why $a_2$ cannot be arbitrarily small.

\medskip

We can now combine the requirements above to obtain a lower bound on $a_2$.  In the small-$a_2$ regime of interest, the characteristic tree-level barrier height for the SFOEWPT is 
set by the height of the intermediate maximum along the $h=0$ direction separating the origin from the singlet minimum.  The intermediate maximum provides an upper bound on the barrier height and 
occurs at
\beq\label{eq:singlet_max}
s(T) \simeq  \frac{-b_3 \pm \sqrt{b_3^2 - 4 b_4 b_2(T)}}{b_4},
\eeq
with the upper (lower) sign again corresponding to $b_3>0$ ($b_3<0$)\footnote{For large enough values of $a_2$, the extremum represented by Eq.~(\ref{eq:singlet_max}) can become a saddle point. However, for values of $a_2$ close to the lower bound we derive, this extremum is indeed a local maximum.}. 
Approximating $\phi_b$ in Eq.~(\ref{eq:tunneling}) with this expression yields the condition
\beq \label{eq:tunneling_bound_2}
\mu^4(T_*) > \frac{b_3^4 \lambda}{48 b_4^3} \left(8-12\xi_*+3\xi_*^2+8 \frac{1+\Delta}{1-\Delta}\left(1-\xi_*\right)^{\frac{3}{2}}\right).
\eeq
Now we can combine Eq~(\ref{eq:tunneling_bound_2}) with Eq.~(\ref{eq:s_stable2}) to obtain an inequality relating $a_2$ and $b_4$:
\beq\label{eq:a2_bound2}
a_2^2 > \frac{4b_4 \lambda}{3 \left(1+\sqrt{1-\xi_*}\right)^4}\left(8-12\xi_*+3\xi_*^2+8 \frac{1+\Delta}{1-\Delta}\left(1-\xi_*\right)^{\frac{3}{2}}\right).
\eeq
This inequality is more stringent than Eq.~(\ref{eq:a2_bound1}), and importantly the RHS does not vanish as $\xi_*\to 8/9$ and the phase transition temperature nears $T_{\rm EW}$. 

Now, for fixed values of $a_2$ and $b_4$,
Eq.~(\ref{eq:b3_lower}) defines a minimum possible value for $|b_3|$ :
\beq 
\left|b_{3,{\rm min}}\right| =\sqrt{\frac{9}{4}\, b_4 (2 m_1^2 - a_2 v^2 + 2 T_{\rm EW}^2 \beta)}.
\eeq 
Taking $\left|b_3\right|\to\left|b_{3,{\rm min}}\right|$ also results in taking $\xi\to 8/9$. 
The RHS of Eq.~(\ref{eq:a2_bound2}) is a monotonically decreasing function of $\xi$, so we can take $b_3 \to b_{3,{\rm min}}$,  $\xi\to 8/9$ in Eq.~(\ref{eq:a2_bound2}) to obtain
\beq
\label{eq:a2_bound3}
a_2^2 > \frac{b_4 \lambda}{4} \frac{\Delta}{1-\Delta}.
\eeq  
This inequality implies $b_4 \sim \mathcal{O}(a_2)$, and thus that both couplings are $ \ll 1$ in the regime of interest. 

We can separately use Eq.~(\ref{eq:tunneling_bound_2}) to obtain a lower bound on $b_4$.  Since $\mu^2(T_*)<\mu^2$, and the RHS of Eq.~(\ref{eq:tunneling_bound_2}) is also a monotonically decreasing function of $\xi$,  we can take 
$\mu^2(T_*)\to \mu^2$ and $b_3 \to b_{3,{\rm min}}$ in Eq.~(\ref{eq:tunneling_bound_2}) 
to obtain the inequality
\beq \label{eq:b4bound_1}
b_4> \frac{\left(2m_1^2 - a_2 v^2 + 2 T_{\rm EW}^2\beta\right)^2 \Delta}{16 \lambda v^4 \left(1-\Delta\right)}.
\eeq
To leading order in $a_2\ll1$ and $b_4\ll1 $, Eq.~(\ref{eq:b4bound_1}) yields 
\beq\label{eq:b4bound_final}
b_4 \gtrsim \frac{m_1^4 \Delta}{4\lambda v^4 \left(1-\Delta\right)},
\eeq
indicating that the singlet quartic cannot be arbitrarily small and allow for successful tunneling. 
Using this lower bound on $b_4$ in Eq.~(\ref{eq:a2_bound3}) and simplifying
gives the desired lower bound on $a_2$:
\beq
\boxed{
\label{eq:a2_bound_final}
a_2 \gtrsim  \frac{m_1^2 }{4 v^2} \frac{\Delta}{1-\Delta} \quad({\rm non-}Z_2).
}
\eeq
One can show that points saturating Eqs.~(\ref{eq:b4bound_final}) and~(\ref{eq:a2_bound_final}) also satisfy the $T=0$ vacuum stability constraint, Eq.~(\ref{eq:s_stable}). 
In summary we find that successful completion of the phase transition is the dominant factor in determining the lower boundary of this parameter space for successful SFOEWPT; the conditions on the depth of minima at zero and finite temperature, Eqs.~(\ref{eq:s_stable}) and~(\ref{eq:higher_smin}), do not by themselves lead to a nontrivial lower bound on $a_2$.
This analytical observation is borne out by the results of our numerical scans with and without imposing the tunneling requirement.

 We again emphasize that the number $\Delta$ simply parameterizes departure from the thin-wall regime and does not reflect the full tunneling requirement, which is implemented in our numerical scans (see Sec.~\ref{sec:numerics}). Our scans find that at small values of $a_2$, such that the barrier height is indeed set by the singlet maximum, points that realize successful tunneling are characterized by $\Delta\gtrsim 0.6-0.8$. Consequently, we will take the intermediate value $\Delta=0.7$ when comparing to the results of our scans, as points with significantly smaller values of $\Delta$ are empirically unlikely to allow for successful tunneling. The fact that the scan results match up well with the semi-analytic bound in Eq.~(\ref{eq:a2_bound_final}) is a non-trivial check of our arguments. Regardless of the precise value of $\Delta$, Eq.~(\ref{eq:a2_bound_final}) indicates that even away from the $Z_2$ limit, the cross-quartic coupling $a_2$ cannot be made arbitrarily small and remain consistent with a viable SFOEWPT, at least for the small mixing angles and approximate treatment of $V_{\rm eff}$ we consider. Note also that as one demands larger and larger values of $\Delta \to 1$, the bound in Eq.~(\ref{eq:a2_bound_final}) suggests that larger and larger values of $a_2$ are required, and at some point the expansion in $a_2\ll 1$ will break down and Eq.~(\ref{eq:a2_bound_final}) should no longer be used. However, with $\Delta\sim 0.7$ as suggested by our scans, this is not an issue.

\begin{figure}[!t]
\begin{center}
\includegraphics[width=0.49\textwidth]{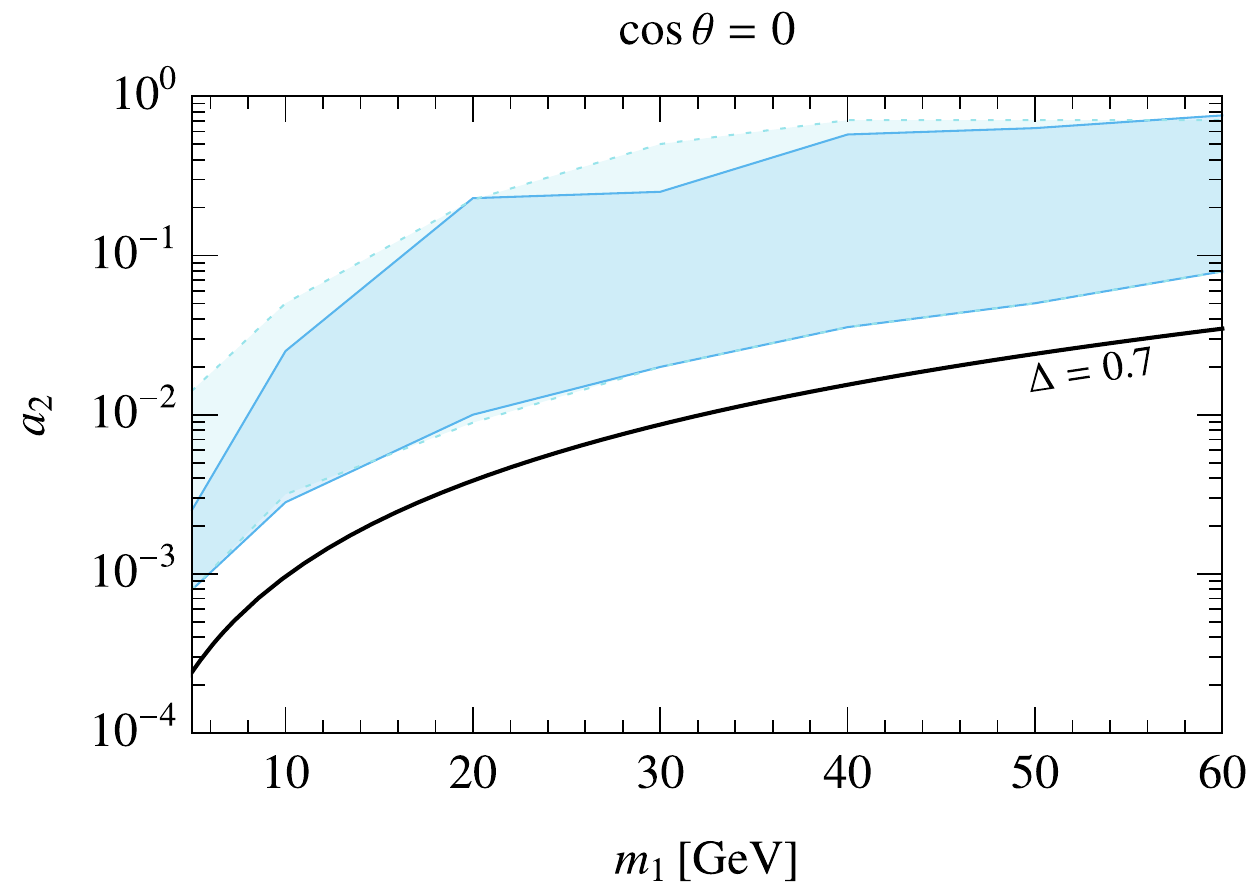}
\includegraphics[width=0.49\textwidth]{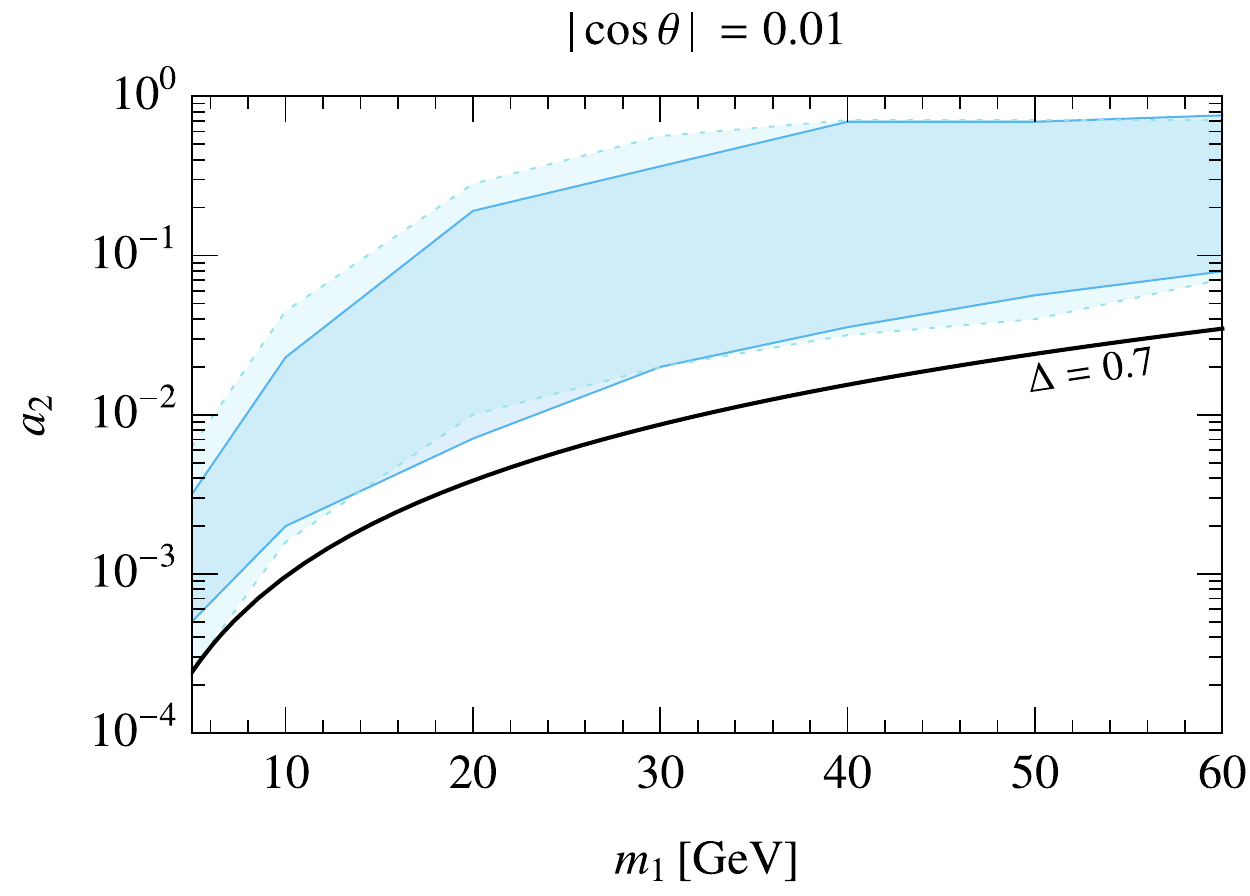}
\caption{Values of the coupling $a_2$ consistent with a strongly first-order electroweak phase transition away from the $Z_2$ limit for $\cos\theta=0$ (left) and $\left|\cos\theta\right|=0.01$ (right). The black curves correspond to the approximate lower bound Eq.~(\ref{eq:a2_bound_final}) with $\Delta=0.7$. The darker shaded region shows the results from a numerical scan of the parameter space using the high-temperature effective potential to $\mathcal{O}(g^2)$ and including a calculation of the tunneling rate, as discussed in Sec.~\ref{sec:numerics}. The lighter shaded region bounded by the dotted contours indicates the results of numerical scans including the leading $\mathcal{O}(g^3)$ corrections in Landau gauge.
}
\label{fig:res_general}
\end{center}
\end{figure}

The semi-analytical lower bounds on $a_2$ derived here for the general potential rely on our approximate treatment of $V_{\rm eff}$ (expanding in small mixing angle, dropping the $S$ linear term, working to $\mathcal{O}(g^2)$) and the tunneling requirement, but we find it to be consistent with the results of numerical parameter space scans in which these approximations are not made, as demonstrated in  Figs.~\ref{fig:res_Z2}-\ref{fig:res_general}.  The blue bands in Figs.~\ref{fig:res_Z2}-\ref{fig:res_general} show the results of numerical scans of the light scalar parameter space, discussed in detail below, together with the semi-analytical lower bounds on $a_2$. 
Fig.~\ref{fig:res_Z2} shows the prediction of Eq.~(\ref{eq:a2_bound_Z2}) for the $Z_2$ limit, and Fig.~\ref{fig:res_general} shows the semi-analytic lower bound of Eq.~(\ref{eq:a2_bound_final}) for $\Delta = 0.7$. Note that the $Z_2$ parameter space will also have an analogous tunneling requirement that must be satisfied in addition to Eq.~(\ref{eq:a2_bound_Z2}), which does not contain any information about the depth of the different minima or the strength of the phase transition. 

Beyond the requirements used in deriving Eqs.~(\ref{eq:a2_bound_Z2}) and~(\ref{eq:a2_bound_final}), other criteria can further reduce the allowed parameter space. In particular, there are additional minima of the potential with both $s$ and $v\neq 0$ that should not be energetically favored over the physical EW vacuum at zero temperature. Furthermore, our analysis thus far has not made use of the SFOEWPT requirement, Eq.~(\ref{eq:sfocond}).  We implement these additional criteria directly in our numerical scans below.

\subsection{Flat directions}
\label{sec:flat}

Up until now, our arguments have focused on two-step transitions in which a barrier is generated between the phases at tree-level. However, there is another possible scenario giving rise to a SFOEWPT without a tree-level barrier: if a flat direction exists in the effective potential at finite temperature to $\mathcal{O}(g^2)$, the $\mathcal{O}(g^3)$ corrections can provide the barrier\footnote{There is in fact an additional possibility, namely that a barrier arises from $T=0$ quantum corrections. However, this occurs at $\mathcal{O}(g^4)$ and requires large couplings of the singlet to the Higgs~\cite{Curtin:2014jma,Chung:2012vg}. Such scenarios are therefore not of interest for transitions with $a_2\ll 1$.}. This corresponds to the breakdown of our power-counting arguments along the flat direction, since a cancellation among the $\mathcal{O}(g^2)$ terms can make formally higher-order corrections more important. This mechanism is emphasized in e.g.~\cite{Espinosa:2011ax} and could in principle provide an exception to our arguments above. However, one can readily show that in the small-$|\cos\theta|$ limit, scenarios with approximate flat directions require larger values of $a_2$
than the lower bounds in Eqs.~(\ref{eq:a2_bound_Z2}) and~(\ref{eq:a2_bound_final}), as we will now demonstrate.

To see this, 
we follow Ref.~\cite{Espinosa:2011ax} in defining two quadratic functions of $s$, $D^2_{h,s}(s)$, such that
\beq
\left. \frac{\partial V_{\rm eff}}{\partial h}\right|_{h^2=D^2_h(s)}=0, \quad \left. \frac{\partial V_{\rm eff}}{\partial s}\right|_{h^2=D^2_s(s)}=0 .
\eeq
A flat direction occurs when these two functions of $s$ coincide,
\beq\label{eq:flat}
D^2_h(s)\simeq D^2_s(s) ,
\eeq
around the critical temperature. Considering the effective potential to $\mathcal{O}(g^2)$ and dropping terms proportional to $\theta^{\prime}$, it is straightforward to see that Eq.~(\ref{eq:flat}) requires $b_3\simeq 0$, $a_2^2\simeq 4 b_4 \lambda$ and
\beq \label{eq:flat_2}
\left|a_2\right| \simeq  \frac{6 m_1^2 + \frac{3}{2} b_4 T_c^2}{\left|3v^2-T_c^2\left(1+3 \frac{v_c^2}{T_c^2} \right)  \right|} >\frac{6 m_1^2}{\left|3v^2-T_c^2\left(1+3 \frac{v_c^2}{T_c^2} \right)\right|}.
\eeq
Note that in contrast to the scenario where the barrier is generated at tree-level,  here $a_2$ can in principle take on either sign. However, only its absolute value is relevant for determining $\lambda_{211}$ for small mixing angles. Using the fact that $v>v_c$ and $v_c/T_c\geq 1$ for a SFOEWPT, the denominator on the RHS of Eq.~(\ref{eq:flat_2}) cannot be larger than $3v^2$, leading to the lower bound
\beq
\left|a_2\right|\gtrsim \frac{2 m_1^2}{v^2}\ ,
\eeq
which is precisely the $Z_2$-symmetric lower bound in Eq.~(\ref{eq:a2_bound_Z2}), extended to allow for negative $a_2$. Therefore, for small values of $|\cos\theta|$, flat directions are unlikely to open up any additional parameter space for a SFOEWPT with smaller $\left|a_2\right|$ than the lower bounds of the previous subsection. Furthermore, one would have to impose additional requirements analogous to those of Sec.~\ref{sec:analytic} to ensure a viable SFOEWPT, which could further reduce the available parameter space. Such scenarios are also more susceptible to the effects of daisy resummation, which tends to reduce the height of the barrier, and $\mathcal{O}(g^4)$ Coleman-Weinberg corrections, which can spoil the requisite flatness of the potential.

Away from the $|\cos\theta|\to 0$ limit that we have been considering, flat directions can play a more important role, extending the viable SFOEWPT parameter space down to smaller $|a_2|$ than required for a SFOEWPT arising from tree-level effects. 
In these cases, one still expects a concrete lower bound on $\left|\lambda_{211}\right|$. However, since we are restricting ourselves in the present work to the experimentally challenging small-$|\cos\theta|$ limit, we do not comment further on this possibility here.

\subsection{Numerical Analysis}
\label{sec:numerics}

The semi-analytic arguments of the previous subsections made use of a simplified treatment of the finite-temperature effective potential, dropping $\cos \theta$-dependent and singlet tadpole terms, as well as an approximate tunneling criterion. In this subsection, we detail our numerical scans of the parameter space, which find points featuring a strong first-order electroweak phase transition without these approximations and assumptions. 

The method for determining the parameter space consistent with a strong first-order electroweak phase transition is that utilized in Ref.~\cite{Chen:2017qcz}, to which we refer the reader for further details. For a given $m_1$ and $\cos \theta$, we scan logarithmically in $a_2, b_3/v \in [10^{-4}, 1]$, and $b_4 \in [10^{-5},1]$ (for $m_1=5$ GeV we extend our $b_4$ scans down to $10^{-6.5}$). We demand that the electroweak vacuum is the global minimum of the tree-level potential. Starting at $T=0$, we minimize the potential and scan up in temperature until the deepest minimum found by the minimization routine corresponds to a phase with restored EW symmetry. 
 If the $h=0$ phase coexists with the broken phase for some range of temperatures, we compute $T_c$ by finding the temperature that minimizes the difference between the free energy of the two phases. 
The code finds the highest temperature transition at which EW symmetry is spontaneously broken, and allows for transitions into vacua with non-vanishing expectation values for both scalars.
This prescription specifically searches for phase transitions at which electroweak symmetry is spontaneously broken.   Transitions along other directions could also be interesting from the standpoint of e.g.~gravitational wave generation, but we do not consider them further here.

Points for which we find $v_c/T_c\geq 1$ are then passed to a tunneling routine to verify that the phase transition in fact completes. We use the \texttt{CosmoTransitions} package~\cite{Wainwright:2011kj} to find the phase transition completion temperature, $T_*$, defined as the temperature at which a fraction $1/e$ of the Universe is found remaining in the symmetric phase. In practice, this amounts to varying the temperature, solving the 3D Euclidean equations of motion for the thermal ``bounce'' (see~\cite{Quiros:1999jp, Wainwright:2011kj}) using \texttt{CosmoTransitions}, computing the 3D Euclidean action, $S_3$, corresponding to the bounce solution, and checking whether 
\beq\label{eq:nucl}
\frac{S_3}{T}&\simeq& 117 - 4 \log \left(\frac{T}{100 \, {\rm GeV}}\right) - 4\log\left(\frac{\beta/H}{100}\right).
\eeq
Here $\beta/H(T) \simeq T d(S_3/T)/dT$ and we have assumed a weak-scale transition with fast-moving bubble walls~\cite{Kozaczuk:2015owa} (see~\cite{Enqvist:1991xw, Gould:2019qek, Caprini:2019egz} for a more detailed explanation). The temperature $T_*$ satisfying the above equality is the temperature at which the PT completes\footnote{Note that we require thermal transitions, and do not consider those for which the $O(4)$-invariant bounce is the lowest-action configuration.}.

It is possible for the bounce action $S_3/T$ to never become small enough to allow any value of $T$ to satisfy Eq.~(\ref{eq:nucl}). In that case, percolation does not occur and the parameter point is excluded from our scans. It is also possible for the numerical tunneling algorithm to fail, in which case again the point is excluded from our results. Alternatively, the phase transition can be supercooled into a period of vacuum energy domination. In other words, $T_*$ defined by Eq.~(\ref{eq:nucl}) can be so low that the radiation energy density drops below the vacuum energy in the symmetric phase. Eq.~(\ref{eq:nucl}) only applies in a radiation-dominated universe, so if supercooling to an inflationary phase is predicted, the criterion for PT completion must be modified. In this case one finds that the phase transition is generally unlikely to complete~\cite{Ellis:2018mja}. We therefore also require $T_*$ to satisfy
\beq
\frac{\pi^2}{30} g_*(T_*)T_*^4 > \Delta V
\eeq
where $\Delta V$ is the difference in vacuum energy density (\emph{not} free energy density) between the symmetric and broken phases and $g_*(T)$ is the effective number of relativistic degrees of freedom at $T$.

The results of our numerical scans, imposing all of the requirements discussed above, are shown in Figs.~\ref{fig:res_Z2}-\ref{fig:res_general} for the $Z_2$ and non-$Z_2$ cases, respectively. For the latter we take $\left|\cos\theta\right|=0, 0.01$ as representative values, although we do not expect the predictions to change significantly as long as $|\cos\theta|$ is small. In these figures, the shaded regions indicate values of $a_2$ for which our numerical analysis finds a strong first-order EWPT satisfying all of the  above requirements. The scans reflected in dark blue utilize the finite temperature effective potential up to $\mathcal{O}(g^2)$ with a high-$T$ expansion, while the lighter shaded regions show the results of scans retaining the leading $\mathcal{O}(g^3)$ corrections from the electroweak gauge bosons. The results support the conclusion that $a_2$ cannot be arbitrarily small and still be consistent with a cosmologically viable SFOEWPT. 

Away from the $Z_2$ limit, the numerical scans do not quite access values of $a_2$ saturating the approximate bound in Eq.~(\ref{eq:a2_bound_final}). We suspect that this is due to a combination of our approximate treatment of the tunneling criterion in Eq.~(\ref{eq:tunneling}) and the fact that the bound in Eq.~(\ref{eq:a2_bound_final}) is only saturated for very specific values of the couplings that can require a high degree of tuning to achieve and therefore be missed by our scans, which have a finite spacing.

In all cases, the analytic and numerical lower bounds result in large enough minimum values of $a_2$ that the small-mixing expansion for the coupling $\lambda_{211}$, Eq.~(\ref{eq:L211}), can be used.  Also, we have also performed some extended scans including negative values of $a_2$ along with the leading $\mathcal{O}(g^3)$ finite-$T$ corrections. We did not find additional parameter space accommodating a SFOEWPT, which is consistent with our power counting arguments and the discussion about flat directions in Sec.~\ref{sec:flat}. We therefore take the results of Figs.~\ref{fig:res_Z2}-\ref{fig:res_general} as indicative of the typical range of $a_2$ (and hence $\lambda_{211}$) values required for a SFOEWPT with $|\cos\theta|\ll 1$. The resulting implications for testing the EWPT with exotic Higgs decays are the subject of the next section.

\section{Implications for exotic Higgs decays}
\label{sec:brexo}

We have argued in the previous section that requiring a strong first-order EWPT induced by a new light scalar suggests a lower bound on the Higgs portal coupling, $a_2 S^2 \left|H \right|^2$, which becomes independent of the Higgs-singlet mixing angle in the small-mixing limit preferred by experiment. Using Eqs.~(\ref{eq:decay}) and~(\ref{eq:L211}), we can directly translate this lower bound on the portal coupling into a lower bound on the branching ratio of the SM-like Higgs into two singlet-like scalars, $h_2\to h_1h_1$ when the decay is kinematically allowed.
As the SM-like Higgs has an accidentally small width  within the SM,  even relatively small values of $ a_2 \approx \lambda_{211}/v$ can yield experimentally interesting deviations in the resulting Higgs branching fractions, making exotic Higgs decays a leading collider probe of light SM-singlet degrees of freedom~\cite{Curtin:2013fra}.  

In this section we analyze the extent to which the LHC and future colliders can probe the EWPT through searches for exotic Higgs decays. 
The specific signatures of and experimental sensitivity to  $h_2\to h_1 h_1$ decays depends on the properties of the singlet-like scalar, $h_1$. Here we consider both visible and invisible Higgs decays.

\subsection{Visible (prompt) decays} 
First we consider the case in which $h_2$ decays visibly and promptly. This requires mixing between the light scalar and the Higgs, i.e.,~$\cos\theta \neq 0$, in which case the singlet-like state decays through its small mixing with the Higgs to SM final states. 
For $m_1\gtrsim 2 m_b$, $h_1$ is generically short-lived on collider scales, i.e.~has proper decay length $< 0.1$ mm: above the $b$ threshold, obtaining proper decay lengths $c\tau_s > 0.1$ mm requires $|\cos\theta|$ to be extremely small, $ < 10^{-4}$ (for $m_b < m_1<2m_b$,  $c\tau_s > 0.1$ mm requires $|\cos\theta| \lesssim 10^{-3.5} $). Accordingly, we consider prompt decays.  It is worth noting that searches targeting prompt decays should generally also have good reach for $h_1$ with small but measurable proper lifetimes, as demonstrated for the $4b$ channel in Ref.~\cite{Aaboud:2018iil}.

\begin{figure}[!t]
\begin{center}
\includegraphics[width=0.6\textwidth]{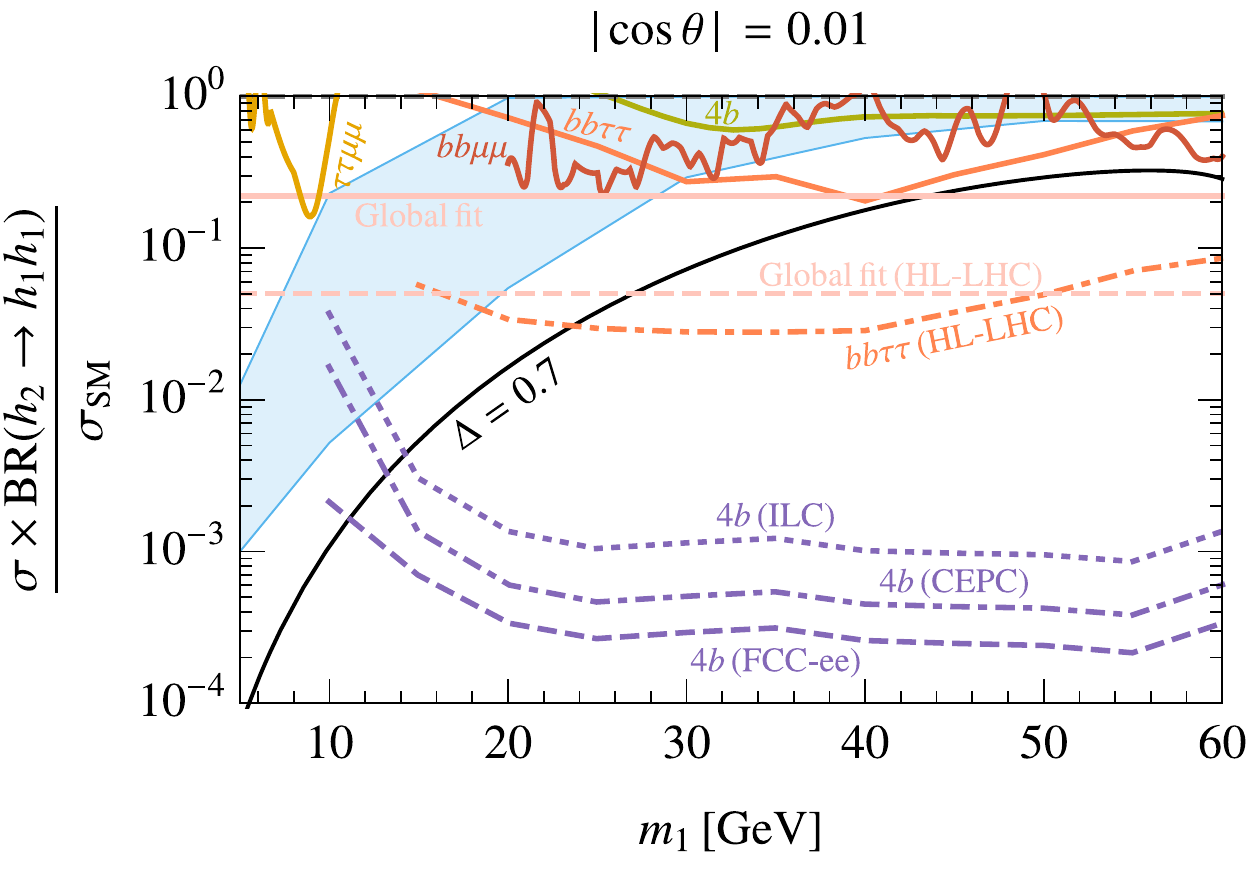}

\caption{Exotic Higgs decays as a probe of strongly first-order electroweak phase transition induced by a light scalar with mass $m_1$. The various labeled contours show the current and future sensitivity to $h_2\to h_1h_1$ assuming prompt visible decays.  Leading current (solid) and projected  (dashed) LHC sensitivities are shown in the $\tau\tau\mu\mu$ (gold) \cite{Aad:2015oqa,Sirunyan:2019gou}, $bb\mu\mu$ (red)
 \cite{Aaboud:2018esj,Sirunyan:2018mot}, $bb\tau\tau$ (orange) \cite{Sirunyan:2018pzn, CMS:2019rsy}, and
  $4b$ (green) \cite{Aaboud:2018iil} channels. For $bb\mu\mu$ and $\tau\tau\mu\mu$ limits, the stronger of the two ATLAS  and CMS  bounds is shown at any given mass point. The  indirect limit on the total exotic branching fraction from \cite{Aad:2019mbh} is indicated by the pink line, and the HL-LHC projection for the same quantity \cite{Cepeda:2019klc} is shown with the pink dashed line.  Also shown in purple are the projected future sensitivities from $h_2\to h_1h_1\to 4b$ searches at the ILC (dotted), CEPC (dot-dashed), and FCC-ee (dashed) \cite{Liu:2016zki}.   The black curve corresponds to the approximate lower bound in Eq.~(\ref{eq:a2_bound_final}) with $\Delta=0.7$. The blue shaded region shows points predicting a strong first-order EWPT with successful tunneling obtained from numerical scans. }
\label{fig:visiblereach}
\end{center}
\end{figure}

Visible, prompt $h_2\to h_1h_1$ decays are dominated by low-mass hadronic states, and are thus particularly challenging signals for the LHC.  Nonetheless, many LHC results are now sensitive to this signal.  The most important direct LHC limits on $h_2\to 2 h_1$ for $m_1>15$ GeV are currently from searches for $bb\mu\mu$
 \cite{Aaboud:2018esj,Sirunyan:2018mot}, $bb\tau\tau$ \cite{Sirunyan:2018pzn}, and
 secondarily $4b$ \cite{Aaboud:2018iil}.  In the mass range $5\,\mathrm{GeV} < m_1 < 10$ GeV, CMS' Run II low-mass search \cite{Sirunyan:2019gou} and ATLAS' Run I search \cite{Aad:2015oqa} in the $\tau\tau\mu\mu$ final state set the leading direct constraints.   For most values of $m_{1}$, however, the constraints set by these direct searches are less stringent than the indirect bound provided by global fits to Higgs
 properties \cite{Sirunyan:2018koj, Aad:2019mbh}, which currently gives an upper limit on non-SM decays of
 BR($h_2 \to $ exotic)$<0.21$.  Forecasts for the HL-LHC anticipate an ultimate sensitivity of  BR($h_2 \to $ exotic)$<0.05$ from global fits \cite{Cepeda:2019klc}.
As an exemplar of LHC sensitivity to visible, prompt decays, we consider the CMS forecast for a direct search in the  $bb\tau\tau$ channel at the HL-LHC  \cite{CMS:2019rsy}, which  will attain sensitivities to BR$(h_2\to h_1 h_1) \lesssim 0.03$  over much of the relevant mass range, surpassing the reach of global fits.  These results are summarized in Fig.~\ref{fig:visiblereach}, where we have used the program {\tt HDECAY} \cite{Djouadi:2018xqq} to compute $h_1$ branching ratios to specific final states as a function of $m_1$.

The prospects for visible $h_2\to h_1h_1$ decays are substantially better in the low-background environments provided by proposed electron-positron colliders. Sensitivity to the dominant $h_2\to 4b$ decay is projected to range from  BR$(h_2\to4b)<9\times 10^{-4}$ for the ILC, $4\times 10^{-4}$ for the CEPC, and $3\times 10^{-4}$ for the FCC-ee  for $m_1=30$ GeV \cite{Liu:2016zki}.  To obtain the estimated sensitivity to $h_2\to h_1 h_1$ in Fig.~\ref{fig:visiblereach}, we use these relative sensitivities at $m_1=30$ GeV to rescale the CEPC forecasts of Ref.~\cite{Liu:2016zki} to obtain estimates for the ILC and FCC-ee for other mass values.  Dedicated projections for exotic decays of the SM-like Higgs at future $e^+e^-$ colliders do not exist in the regime $m_1 < 2 m_b$, but a rough 
estimate suggests that much of the interesting parameter space below the $b$ threshold could be probed.

 To study the impact of these searches on scenarios with a strong first-order EWPT catalyzed by a light scalar, in Fig.~\ref{fig:visiblereach} we also show the approximate lower bound 
of Eq.~(\ref{eq:a2_bound_final}) with $\Delta=0.7$ (with $\lambda_{211}$ evaluated in the $\cos\theta\to0$ limit) as well as the range of branching ratios obtained from the numerical scans considering the gauge-invariant $\mathcal{O}(g^2)$ effective potential described in Sec.~\ref{sec:numerics}. Including the leading $\mathcal{O}(g^3)$ corrections yields very similar results, as indicated by Fig.~\ref{fig:res_general}. The scans use $\left|\cos\theta\right|=0.01$, although the EWPT results are insensitive to the value of $|\cos\theta|$ as long as it is small (see Fig.~\ref{fig:res_general}). 
From Fig.~\ref{fig:visiblereach}, we see that the LHC already strongly limits the parameter space for which a light singlet-like scalar can drive the EWPT to be strongly first-order.  All first-order points in our scan for $\left|\cos\theta\right|=0.01$ above $m_1\simeq 28$ GeV are already excluded by bounds on BR($h_2 \to $ exotic) and limits from $4b$, $bb\tau\tau$, and $bb\mu\mu$ searches. The HL-LHC would extend this sensitivity down to the $\sim 18-22$ GeV range. Eq.~(\ref{eq:a2_bound_final}) and our numerical scans also suggest that future electron-positron colliders will be able to probe singlet-catalyzed SFOEWPTs all the way down to the $b\bar{b}$ threshold using searches in the $4b$ channel alone.

Although we only show results for $\left|\cos \theta\right|=0.01$, we have verified that our results are insensitive to the precise value of the mixing angle as long as its magnitude is small and it is large enough for $h_1$ to decay (sufficiently) promptly to have good acceptance in searches for prompt exotic Higgs decays.

\subsection{Invisible decays}  
Another possibility is that the light scalar is stable on collider timescales, as can arise either in the $Z_2$-symmetric scenario or more generally when $\cos\theta \simeq 0$.  
In this case, searches for invisible Higgs decays at the LHC and future colliders have excellent sensitivity to the parameter space in which a light scalar can catalyze a SFOEWPT. 

The leading LHC limit on invisible Higgs decays is currently BR$(h_2\to\met) < 0.22$ at 95\% CL, from the CMS global fit to Higgs measurements  \cite{Sirunyan:2018koj}. For future colliders, direct searches for invisible decays are always more sensitive than the bound on the total exotic width of the Higgs.
  The forecasted sensitivity at the HL-LHC from combined ATLAS and CMS measurements of $h_2\to \met$ alone  is  BR$(h_2\to\met)<0.025$ \cite{Cepeda:2019klc}.  Future $e^+e^-$ machines will again enable more precise measurements, 
 with $95\%$ CL limits on BR$(h_2\to\met)$ forecasted to reach   
$0.0022$ at the 500 GeV ILC, $0.0027$ at the CEPC, and  
$0.0019$ at the FCC-ee running at a center-of-mass energy of 365 GeV  \cite{deBlas:2019rxi}.  Ref.~\cite{deBlas:2019rxi} also forecasts that the ultimate combined power of the FCC-ee/eh/hh program will push the exclusion by another order of magnitude to BR$(h_2\to\met)<0.00024$.

\begin{figure}[!t]
\begin{center}
\includegraphics[width=0.49\textwidth]{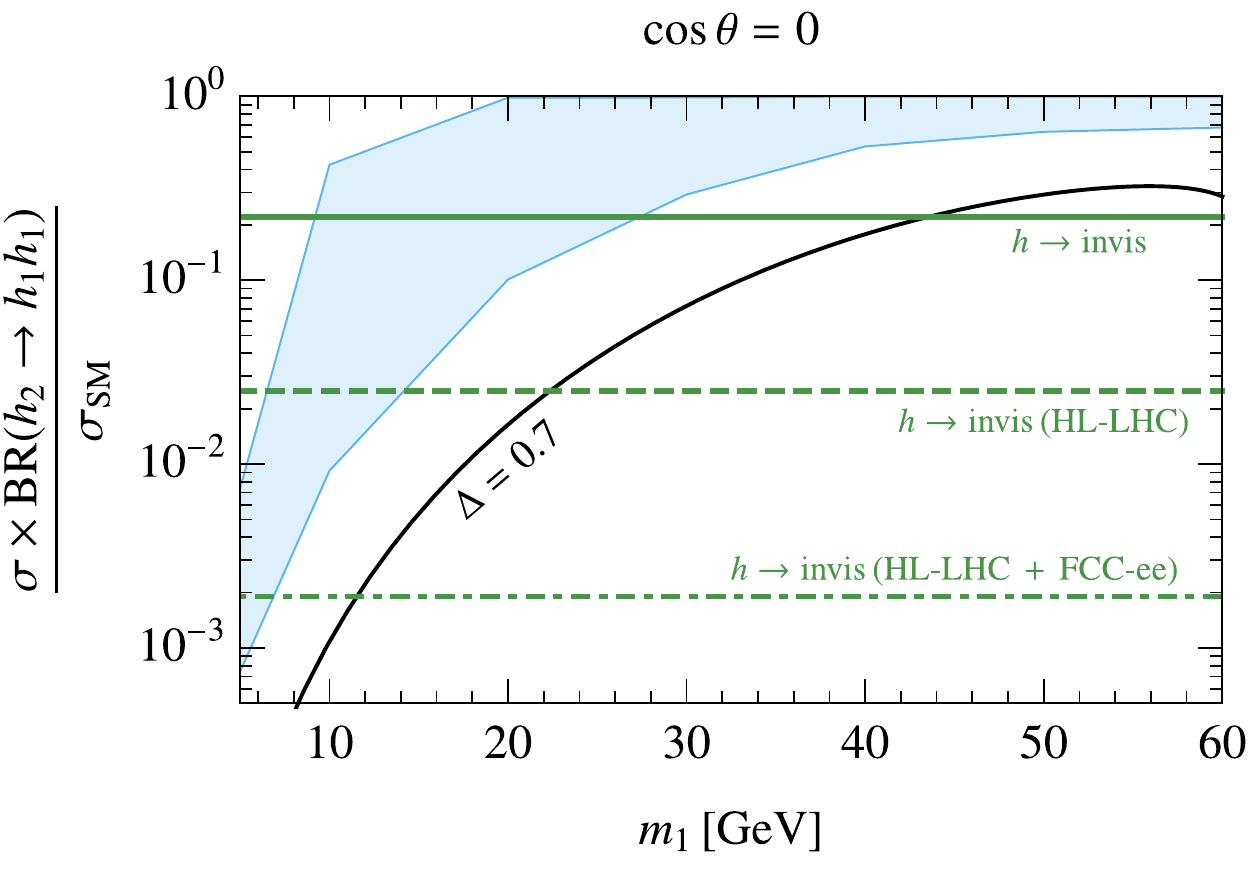}
\includegraphics[width=0.49\textwidth]{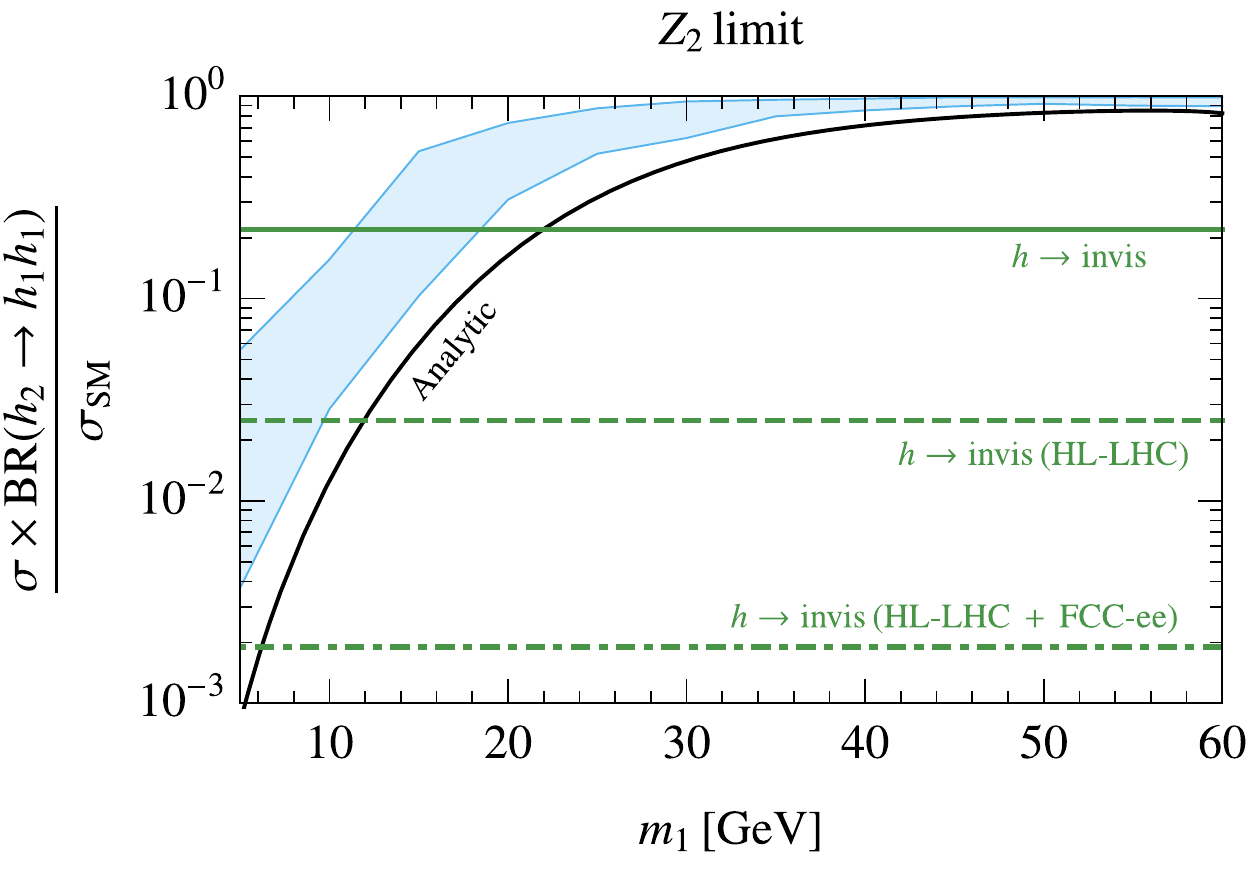}
\caption{ Current and expected future sensitivity to $h_2\to h_1h_1$ decays for invisible $h_1$.  The current exclusion is shown with the horizontal green line  \cite{Sirunyan:2018koj}, with future projections for the HL-LHC \cite{Cepeda:2019klc} and the FCC-ee \cite{deBlas:2019rxi}  
 shown with dashed and dot-dashed lines, respectively. The corresponding ILC and CEPC projections lie close to the FCC-ee prediction (see text). The solid black lines correspond to the (semi-)analytic estimates in Eqs.~(\ref{eq:a2_bound_Z2}) and~(\ref{eq:a2_bound_final}). The blue shaded region shows points predicting a strong first-order EWPT with successful tunneling obtained from numerical scans. }
\label{fig:invisiblereach}
\end{center}
\end{figure}

In Fig.~\ref{fig:invisiblereach} we show current and expected future sensitivities to invisible Higgs decays. The left panel shows the $\cos\theta=0$ limit of the full model, while the right panel shows the more restrictive $Z_2$-symmetric scenario. We also show the (semi-)analytic lower bounds of Eqs.~(\ref{eq:a2_bound_Z2}) and~(\ref{eq:a2_bound_final}), as well as the results of our numerical analysis in the shaded region, again considering the effective potential to $\mathcal{O}(g^2)$. We find that in the $Z_2$ case, existing bounds on the invisible Higgs branching ratio already exclude scenarios in which a light scalar with mass $20\,\mathrm{GeV}\lesssim m_1 < m_h/2$ drives the EWPT to be strongly first-order. In the non-$Z_2$ zero-mixing case, the current constraint sits closer to $m_1\lesssim 30$ GeV from our numerical scans (and $\sim40$ GeV if the semi-analytic bound were saturated). The high-luminosity LHC will be able to exclude cases with $m_1 \gtrsim 5-10$ GeV in the $Z_2$-symmetric limit, while both our numerical scans and semi-analytic arguments suggest that a future Higgs factory such as the FCC-ee will be able to probe a similar range of masses away from the $Z_2$ limit compatible with a SFOEWPT.

\section{Discussion and conclusions} 
\label{sec:conc}

We have explored scenarios in which the interaction between the SM Higgs boson and a light scalar field catalyzes a first order electroweak phase transition, and have demonstrated that there indeed remains viable, albeit potentially fine-tuned, parameter space for SFOEWPTs featuring a new scalar state with mass lighter than half the Higgs mass.  We have provided both semi-analytic and numerical estimates of the range of couplings and resulting exotic Higgs branching ratios consistent with a SFOEWPT in the small Higgs-singlet mixing regime, finding that the requirement of successful completion of the phase transition plays a key role in limiting the viable parameter space.  The experimental prospects for exploring the resulting parameter space in searches for exotic Higgs decays are promising. Our study indicates that light scalars with masses between $\sim 10$ GeV $-m_h/2$ catalyzing a strong first-order EWPT  will be conclusively probed by exotic Higgs decay searches at both the LHC and future lepton colliders in the otherwise challenging regime where the singlet-Higgs mixing angle $|\cos\theta|\lesssim 0.01$. 

Our primary analyses were carried out in an $\mathcal{O}(g^2)$ approximation to the effective potential.  We expect this approximation to be parametrically justified in the general parameter space of interest for the light singlet model studied here, where the barrier between EW-symmetric and EW-breaking minima is generated at tree-level.  We have further verified that additional parameter space does not open up along flat directions, where  cancellations among the terms in the potential at $\mathcal{O}(g^2)$ can cause these power-counting arguments to fail.   This $\mathcal{O}(g^2)$ approach has the major advantage of being manifestly gauge invariant~\cite{Patel:2011th,Katz:2015uja}, and we have also performed scans including the leading $\mathcal{O} (g^3)$ contributions in Landau gauge, finding very similar results for small $a_2$, as expected from power counting arguments.  However we urge the reader to keep in mind several caveats:
\begin{itemize}
\item Our analysis does not take into account zero temperature radiative corrections to the effective potential, which might affect the allowed parameter space for a SFOEWPT, possibly reducing it. The high-temperature expansion for the thermal corrections also does not take into account potentially important effects such as daisy resummation, which again can impact the SFOEWPT parameter space. Even when these additional contributions are included (at the expense of gauge-independence), there can still be large uncertainties inherent in perturbative calculations of phase transition properties due to infrared divergences that arise near the symmetric phase. Ideally, our conclusions should be confirmed on the lattice to ensure that all relevant effects on the EWPT are accounted for. 
\item The results of our numerical analysis are limited by the spacing between points in our scans. It could in principle be possible to find additional (tuned) parameter space not resolved by our analysis, though the good agreement of our semi-analytic arguments with the output of our numerical analysis lends confidence to our delineation of the viable parameter space. \item We have focused on small mixing angles between the new scalar and the Higgs, while currently for certain singlet masses the mixing is allowed to be somewhat larger than the $|\cos\theta|\lesssim\mathcal{O}(0.01)$ values considered. The values of our lower bounds on exotic Higgs branching ratios will change for sufficiently large mixing angles, although such points are also easier to detect through direct and indirect searches at colliders.
\item Our tunneling calculation was performed numerically and, in some instances, the \texttt{CosmoTransitions} algorithm did not converge. Points for which the calculation failed were discarded from our results, but it could be that the PT could still complete. It would be interesting to compare the results from \texttt{CosmoTransitions} against other methods for determining the bounce, which can provide better convergence in some cases~\cite{Athron:2019nbd}. 
\end{itemize}

With these caveats in mind, our results still demonstrate that exotic Higgs decays provide a powerful test of the electroweak phase transition in scenarios with light scalars.   Our numerical analysis suggest that the LHC already excludes light scalars with masses between $\sim 20$ GeV $-\,m_h/2$ ($\sim 30$ GeV $-\,m_h/2$) with a coupling to the Higgs large enough to induce a SFOEWPT in the $Z_2$ (non-$Z_2$) limit.  Although the current LHC sensitivity is dominated by global fits to Higgs properties, direct searches for exotic decays in the $bb\tau\tau$, $bb\mu\mu$, $4b$ and (when $h_1$ is detector-stable) $\met$ channels are already directly probing the territory that realizes SFOEWPTs. For scalar masses $m_1\gtrsim 20$ GeV, future HL-LHC sensitivity is forecasted to be dominated by direct searches.  At lower scalar masses, however,  searches at future Higgs factories are necessary to fully cover the parameter space that realizes SFOEWPTs.  Searches at future lepton colliders in $h_2\to 4b$ or $h_2\to\met$ are projected to extend this sensitivity down to $\sim 10$ GeV.  Below the $b\bar{b}$ threshold, further development of LHC searches in $h_2\to \tau\tau\mu\mu$ ($h_2\to4\tau$) and lepton collider searches in $h\to 4c$ ($h\to cc\tau\tau$, $h_2\to4\tau$) could provide powerful additional handles on the viable parameter space of interest.  These results provide further motivation for extending the present and planned exotic Higgs decay program at the LHC to the prospective $e^+e^-$ colliders under consideration by the high energy physics community.

Looking forward, we studied prompt or invisible Higgs decays, however tiny but non-zero mixing angles, $|\cos\theta|\lesssim 10^{-4}$, would allow the light scalar to decay a macroscopic distance away from the interaction point.   In the absence of an approximate symmetry, such tiny mixing angles require fine-tuning, but LHC prospects for Higgs decays to long-lived light scalars are in general better than for promptly-decaying scalars, as reviewed in \cite{Lee:2018pag}, and the potential HL-LHC coverage of the EWPT-compatible region should correspondingly be significant. Also, while we have focused on scalar masses above $\sim 5$ GeV, it would be interesting and worthwhile to extend our analysis to even lower masses, where production in meson decays becomes possible and displaced $h_1$ decays become more generic.

\section*{Acknowledgments}
We thank Marcela Carena, Zhen Liu, and Yikun Wang for useful discussions of their related work in a related scenario. We also thank Aqil Sajjad for collaboration in the early stages of this project. The work of JK was supported by Department of Energy (DOE) grant DE-SC0019195 and National Science Foundation (NSF) grant PHY-1719642. MJRM was supported in part under U.S Department of Energy contract DE-SC0011095. The work of JS is supported by DOE Early Career grant DE-SC0017840. The authors performed this work in part at the Aspen Center for Physics, which is supported by NSF grant PHY-1607611.

\bibliography{Higgs_decays_EWPT}
\bibliographystyle{JHEP}

\end{document}